\documentclass[a4paper,11pt]{article}
\usepackage{jinstpub} 
\usepackage{lineno}

 \proceeding{7$^{\text{th}}$ International Conference on Micro Pattern Gaseous Detectors\\
 Rehovot, Israel\\
 11–16 December 2022}

\title{\boldmath Impact of the gas choice and the geometry on the breakdown limits in Micromegas detectors}

\author[a,b,1]{P. Gasik,\note{Corresponding author.}}
\author[c]{T. Waldmann}
\author[c]{L. Fabbietti}
\author[c]{T. Klemenz}
\author[c,d]{L. Lautner}
\author[c]{B. Ulukutlu}

\affiliation[a]{GSI Helmholtzzentrum f\"{u}r Schwerionenforschung GmbH (GSI),\\ Darmstadt, Germany}
\affiliation[b]{Facility for Antiproton and Ion Research in Europe GmbH (FAIR),\\ Darmstadt, Germany}
\affiliation[c]{TUM School of Natural Sciences, Technische Universit\"{a}t M\"{u}nchen,\\ Garching, Germany}
\affiliation[d]{European Organization for Nuclear Research (CERN),\\ Geneva, Switzerland}

\emailAdd{p.gasik@gsi.de}

\abstract{In this study we investigate the stability limits of Micromegas detectors
upon irradiation with alpha particles. The results are obtained with meshes with different optical transparency and geometry of wires. The measurements are performed in Ar- and Ne- based mixtures
with different CO$_2$ content. We observe that the breakdown limit strongly depends on the gas and that a higher amount of quencher in the mixture does not necessarily correlate with higher stability. In addition, we observe discharge probability scaling with the wire pitch. This suggests that a Micromegas mesh cell can be treated as an independent amplification unit, similar to a hole in a GEM foil. The outcome of these studies provides valuable input for further optimization of MPGD detectors, multi-layer stacks in particular.}

\keywords{Gaseous detectors, Micropattern gaseous detectors (MSGC, GEM, THGEM, RETHGEM, MHSP, MICROPIC, MICROMEGAS, InGrid, etc),}

\usepackage{tabularx}   
\usepackage{booktabs}   
\usepackage{multirow}
\usepackage{siunitx}
\DeclareSIUnit\clight{\text{\ensuremath{c}}}
\DeclareSIUnit[number-unit-product = ]\percent{\char`\%}

\usepackage{xspace} 

\newcommand{\murm}{%
  \ifmmode
    \mathchoice
        {\hbox{\normalsifze\textmu}}
        {\hbox{\normalsize\textmu}}
        {\hbox{\scriptsize\textmu}}
        {\hbox{\tiny\textmu}}%
  \else
    \textmu
  \fi
}

\newcommand{\pp}{\ensuremath{\mathrm {p\kern-0.05em p}}\xspace}

\newcommand{\figref}[1]{Fig.~\ref{#1}}

\newcommand{\tabref}[1]{Table~\ref{#1}}

\newcommand{\ArCOtwo}{Ar-CO$_2$ (90-10)\xspace}
\newcommand{\ArCOtwoThirty}{Ar-CO$_2$ (70-30)\xspace}

\newcommand{\dsrc}{\ensuremath{d_\mathrm{source}}\xspace}

\newcommand{\eg}{e.\,g.\@\xspace}

\newcommand{\ie}{i.\,e.\@\xspace}

\makeatletter\newcommand*{\etc}{etc\@ifnextchar.{}{.\@}}\makeatother

\begin{document}
\maketitle
\flushbottom

\section{Introduction}
The most common MPGD structures, GEM~\cite{GEM} and MICROMEGAS~\cite{MMG}, can provide high position, time, and energy resolutions, high rate capability, and quite reliable spark protection capabilities. However, experience shows that in real experimental conditions, there is a non-zero probability of spark development. If the charge carrier density in the amplification region reaches the critical charge limit, the avalanche transitions to a streamer that ends up in a spark. The minimum avalanche size necessary to trigger a spark varies between 10$^6\,e$, and 10$^7\,e$ and concerns all MPGD-type detectors (see \eg ~\cite{BachmannSpark, MathisGEM, LautnerTH, Bressan}). 

Systematic studies revealed that sparks usually appear in a narrow region of the amplification gap, where the electric field lines are parallel to each other and no quenching by the electric field reduction is possible. When a streamer reaches the cathode of the amplification structure, a full breakdown is observed. The occurrence of spark discharges in a Micromegas detector, and related critical charge limits, have been studied in numerous experimental and simulation works (see \eg ~\cite{BAY2002162} and~\cite{PROCUREUR2010177}, respectively).

A proper understanding of the fundamental limits of mesh structures may provide important input for designing Micromegas-based detectors. For completeness of our discharge studies, conducted initially with (TH)GEMs and summarised in~\cite{MathisGEM, LautnerTH}, we have launched a dedicated discharge investigation campaign with several types of Micromegas detectors. The main motivation of the studies is to measure stability dependency on the mesh geometry, varying the diameter of its wires ($d_{\mathrm{wire}}$) and the distance between them ($a_{\mathrm{wire}}$), thus varying its optical transparency. From the studies presented in~\cite{ALVIGGI2020162359} a clear dependency on the wire geometry is shown pointing to thinner wires and smaller Micromegas cells having higher stability. The conclusions are based on the argument of amplification field uniformity which is supposed to be better in meshes with thinner wires and smaller cells~\cite{ALVIGGI2020162359}. On the other hand, Finite Elements Method calculations of the electric field obtained with different meshes, discussed in~\cite{Bhattacharya_2020}, show that  the peak-to-average value of the electric field is rather constant at $\sim$3, or even slightly drops for thicker wires and larger cells (although, it increases when for the same wire diameter only the distance $a_{\mathrm{wire}}$ is enlarged). Also, the peak values are higher for thinner wires and increase with a larger wire pitch. From the electric field considerations, assuming discharges develop close to the high-field regions, thicker wires and less dense meshes would be preferable. This is also a conclusion presented in~\cite{Bhattacharya_2020}, supported by the streamer development simulations in meshes with $d_{\mathrm{wire}}=18$\,\textmu m, $a_{\mathrm{wire}}=45$\,\textmu m, and $d_{\mathrm{wire}}=28$\,\textmu m, $a_{\mathrm{wire}}=50$\,\textmu m. This result is in contradiction with the measurements presented in~\cite{ALVIGGI2020162359} where meshes of the same geometry show an opposite stability performance. 

\section{Measurements}
In order to further understand the main factors responsible for the discharge behavior of a mesh-based structure, we have performed measurements where a Micromegas detector was irradiated with highly ionizing alpha particles. All detectors used in this study were produced in the CERN EP-DT-DD MPT laboratory with the Micromegas bulk technology. The available meshes are listed in \tabref{tab:qcrit:mmg}, together with their main geometrical parameters. Due to technological reasons, the detectors were produced with different distances $z_{\mathrm{MMG}}$ between the mesh and readout anode. 

\begin{table}[h]
\centering
\caption{A list of meshes used in the discharge studies. All meshes are woven. The main geometrical parameters are specified: $z_{\mathrm{MMG}}$ - amplification gap, $d_{\mathrm{wire}}$ - wire thickness, $a_{\mathrm{wire}}$ - distance between the wire edges (inner dimensions of a mesh cell), density in lines per inch, optical transparency.}
\begin{tabular}{l *5c}
\toprule

\multirow{2}{*}{Mesh} & $z_{\mathrm{MMG}}$ & $d_{\mathrm{wire}}$ & $a_{\mathrm{wire}}$ & Density & Transparency \\[0.5ex]
 & [\textmu m] & [\textmu m] & [\textmu m] & [LPI] & [\%] \\
\midrule
MMG1  & 128 & 13 & 22 & 730 & 39.5 \\[1ex]
MMG2  & 128 & 15 & 25 & 640 & 39.0 \\[1ex]
MMG3  & 125 & 18 & 45 & 400 & 51.0 \\[1ex]
MMG4  & 200 & 30 & 80 & 230 & 52.0 \\
\bottomrule
\end{tabular}
\label{tab:qcrit:mmg}
\end{table}

\begin{figure}[h]
    \centering
    \includegraphics[width=0.33\linewidth]{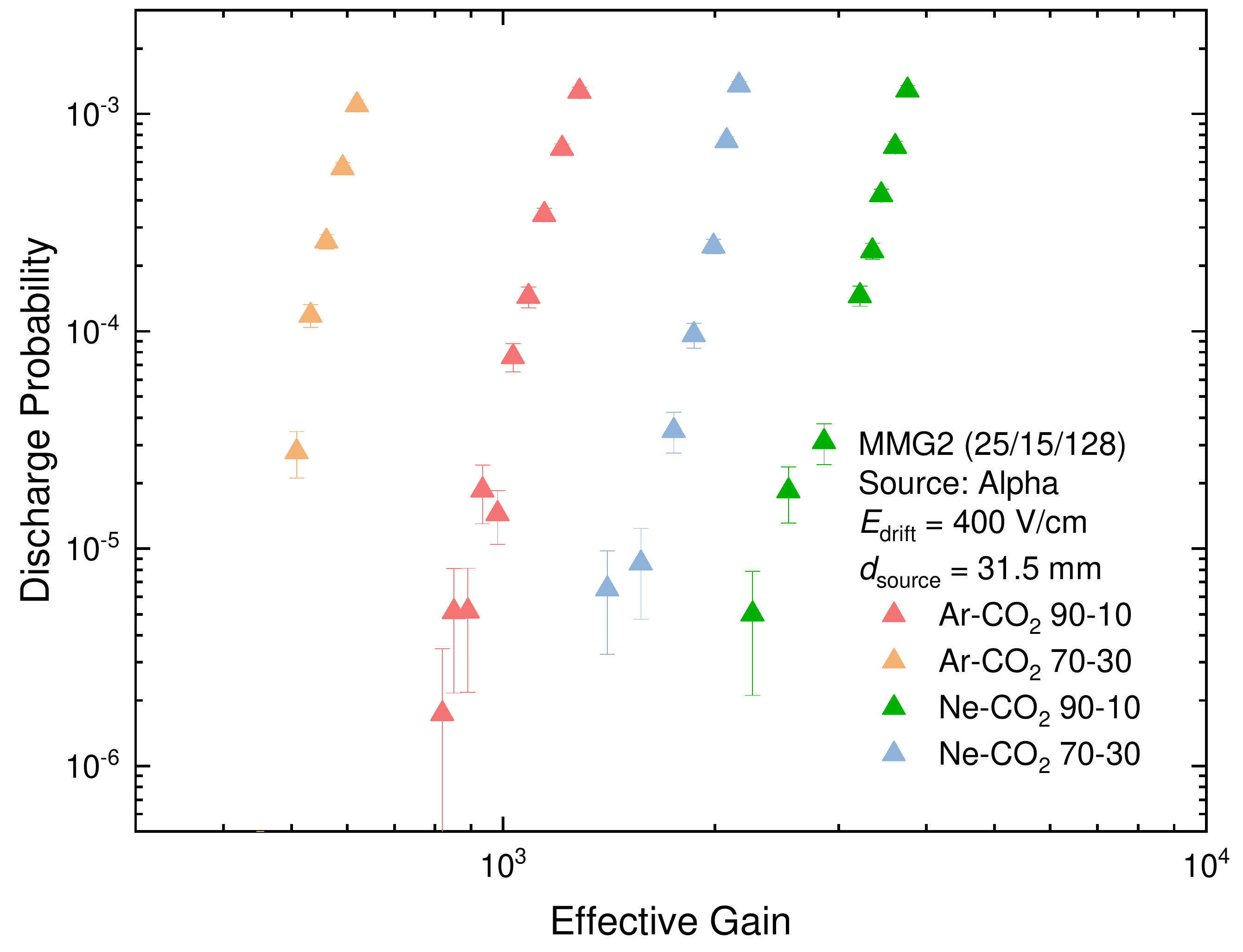}\hfill
    \includegraphics[width=0.33\linewidth]{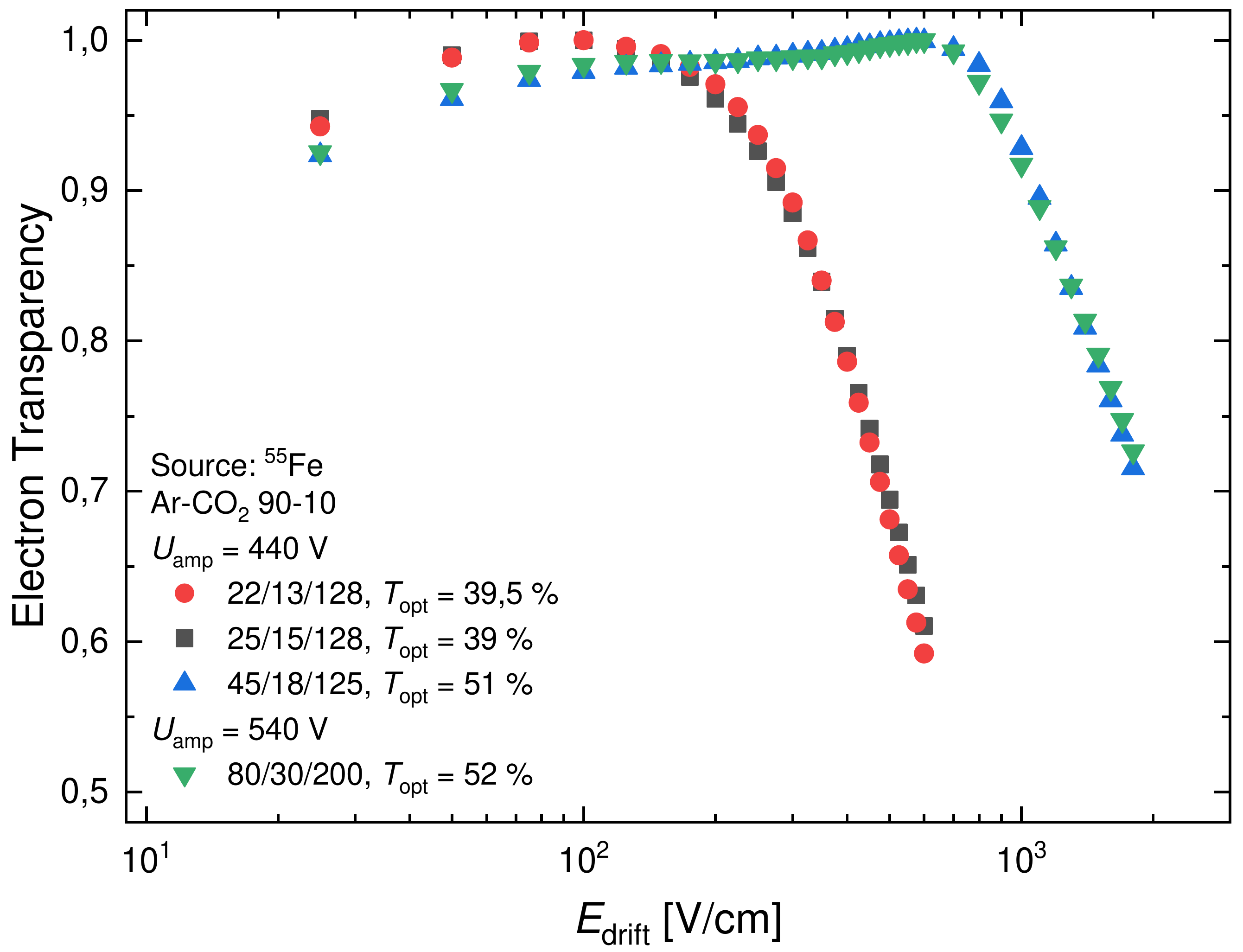}\hfill
    \includegraphics[width=0.33\linewidth]{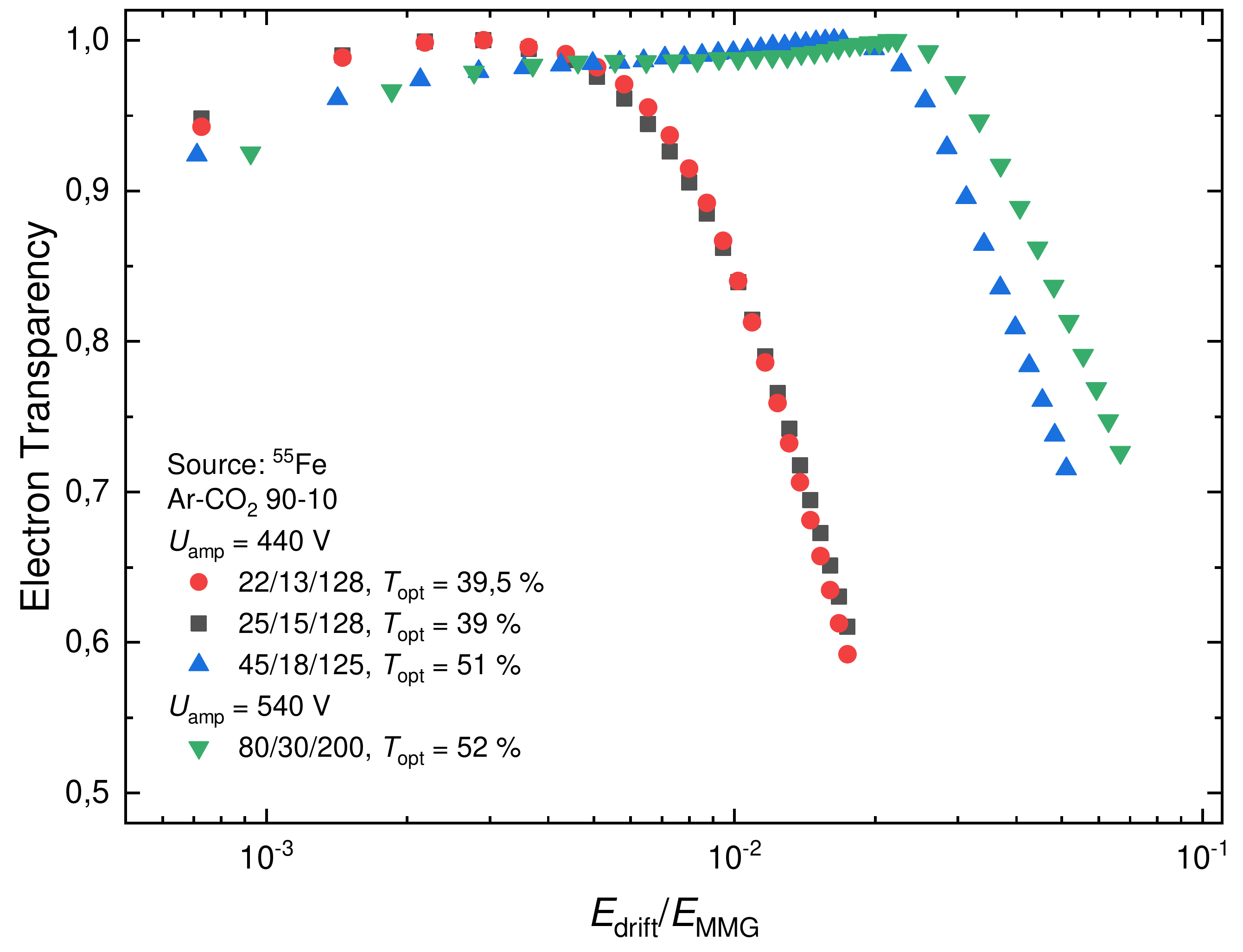}
    \caption{\textit{Left:} discharge probability of the MMG2 structure measured with alpha particles. A clear dependency on the average atomic number of the gas mixture and the quencher content is visible. \textit{Middle and right:} electron transparency of all four meshes measured with an $^{55}$Fe source as a function of drift field (middle) and drift-to-amplification field ratio (right).}
    \label{fig:mmg:transparency}
\end{figure}

In all presented studies the alpha source (mixed nuclide $^{239}$Pu, $^{241}$Am, and $^{244}$Cm) was placed on top of the drift electrode, shooting toward the amplification structure. The left panel in \figref{fig:mmg:transparency} presents the discharge probability per alpha particle measured with the MMG2 structure as a function of its effective gain in four gas mixtures. The effective gain is defined as the ratio of the amplification current induced on the readout plane to the primary current measured in the drift gap. Hence, for electron transparencies of \SI{<100}{\percent} the actual (absolute) gain value is larger than the effective one. A clear gas dependency can be observed in the order of discharge curves, pointing to lighter gases as more stable ones, as observed in the studies with GEMs~\cite{MathisGEM} and THGEMs~\cite{LautnerTH}. For both, Ne- and Ar-based mixtures an anti-correlation can be observed between the stability of the Micromegas and the quencher content of the gas. Similar behavior was observed with three other structures and is in line with previous measurements presented in~\cite{BAY2002162}. The results clearly point to the primary charge density having an essential influence on the probability of spark development. Therefore, one can ask whether mesh cells can be considered independent amplification units, similar to GEM holes. Following the GEM and THGEM comparison~\cite{LautnerTH}, the discharge probability shall scale with the Micromegas cell size, \ie higher discharge rate is expected for large-$a_{\mathrm{wire}}$ meshes (small LPI value) which presumably collect more primary charges in a single cell than denser meshes, where the primary charge cloud is shared among many cells.

It is important to note, however, that the quality and density of the mesh affect multiple aspects of Micromegas performance which should be taken into account in the comparison study. Firstly, high electric fields around thin mesh wires and/or any defects present in the mesh may further reduce sensitivity to the investigated correlations. We believe, however, that highly ionizing alpha particles used in our measurements liberate enough primary ionization to study discharge dependencies at relatively low fields and gains, where the effect of the mesh quality is suppressed. Secondly, the open geometry of a Micromegas structure and poor quenching capabilities of CO$_2$ imply photon and ion feedback issues, especially at high gains. Thus, the measurements in the low-gain region also allow us to neglect this effect. 

Finally, the optical transparency of a mesh also influences its electron transparency. This is taken into account by a detailed characterization of each mesh~\cite{Nikolopoulos_2011, Bhattacharya_2014, Krugerphd}. The electron transparency is measured
with an $^{55}$Fe source as a function of the drift field and the drift-to-amplification field ratio, as shown in \figref{fig:mmg:transparency}. The amplification field $E_{\mathrm{MMG}}$ is approximated by $U_{\mathrm{MMG}}/z_{\mathrm{MMG}}$, where $U_{\mathrm{MMG}}$ is the potential applied to the mesh. The transparency is evaluated by measuring the amplification current while keeping $U_{\mathrm{MMG}}$ constant and normalizing it to the maximum measured current value, where a transparency of \SI{100}{\percent} can be reliably assumed. A clear correlation with the optical transparency is observed, as expected. The maximum transmission for the low-transparency meshes (\SI{\sim39}{\percent}) is observed for a drift field of \SI{\sim100}{\volt\per\centi\meter}, whereas for meshes with an optical transparency of \SI{\sim50}{\percent} the corresponding drift field is $E_{\mathrm{drift}}\approx600$\,\si{\volt\per\centi\meter}. For higher $E_{\mathrm{drift}}$ values the collection efficiency drops as more electric field lines, that originate from the drift cathode, end on the metallic mesh. Thus, for the comparison studies of the four meshes and their absolute gain determination the drift field value of \SI{150}{\volt\per\centi\meter} is chosen, for which the transparency of \SI{\sim100}{\percent} can be safely assumed. Given the $E_{\mathrm{drift}}/E_{\mathrm{MMG}}$ dependency, shown in the right panel of \figref{fig:mmg:transparency}, the assumption holds for the investigated $U_{\mathrm{MMG}}$ voltage range of 400--600\,V.

\begin{figure}[h]
    \centering
    \includegraphics[width=0.33\linewidth]{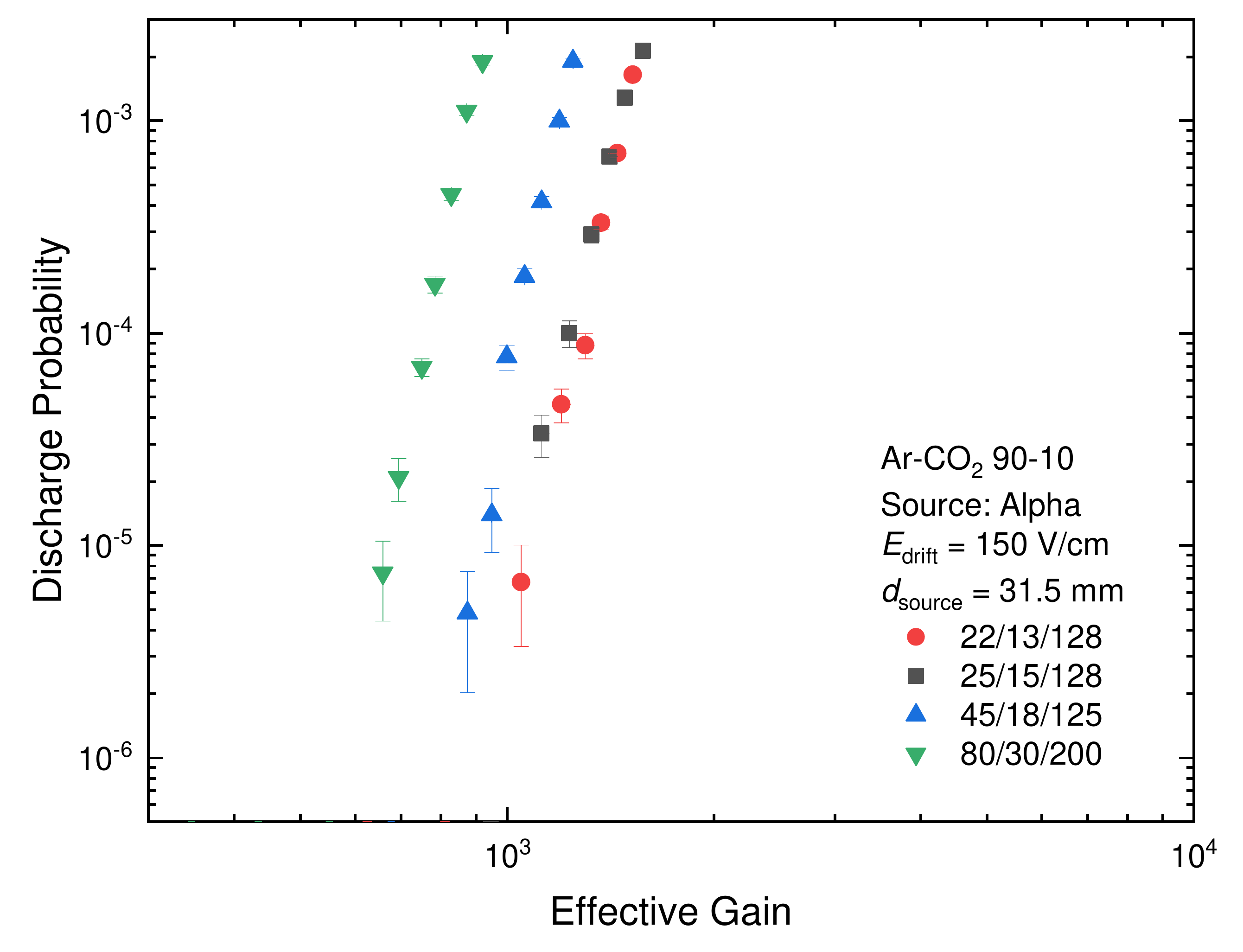}\hfill
    \includegraphics[width=0.33\linewidth]{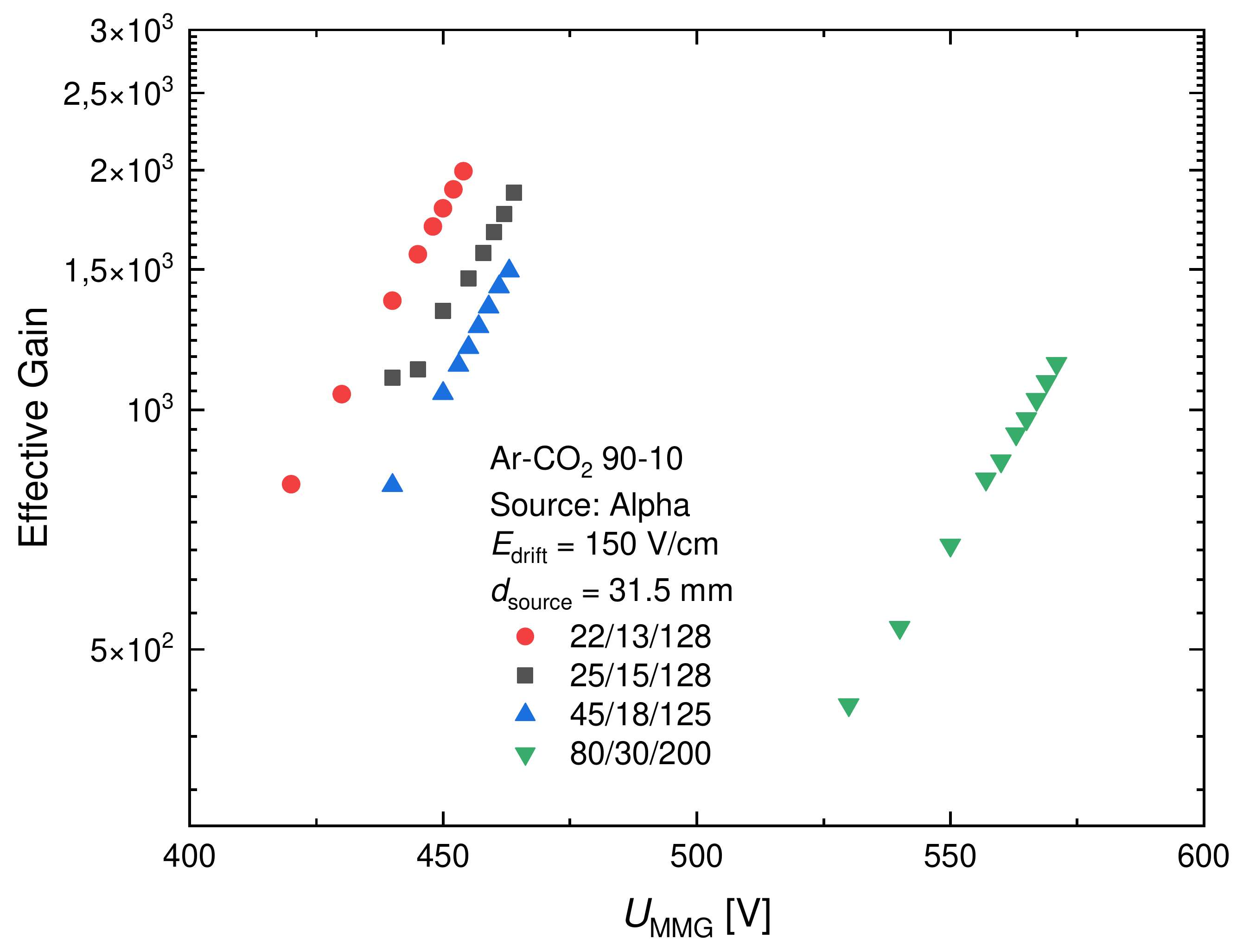}\hfill
    \includegraphics[width=0.33\linewidth]{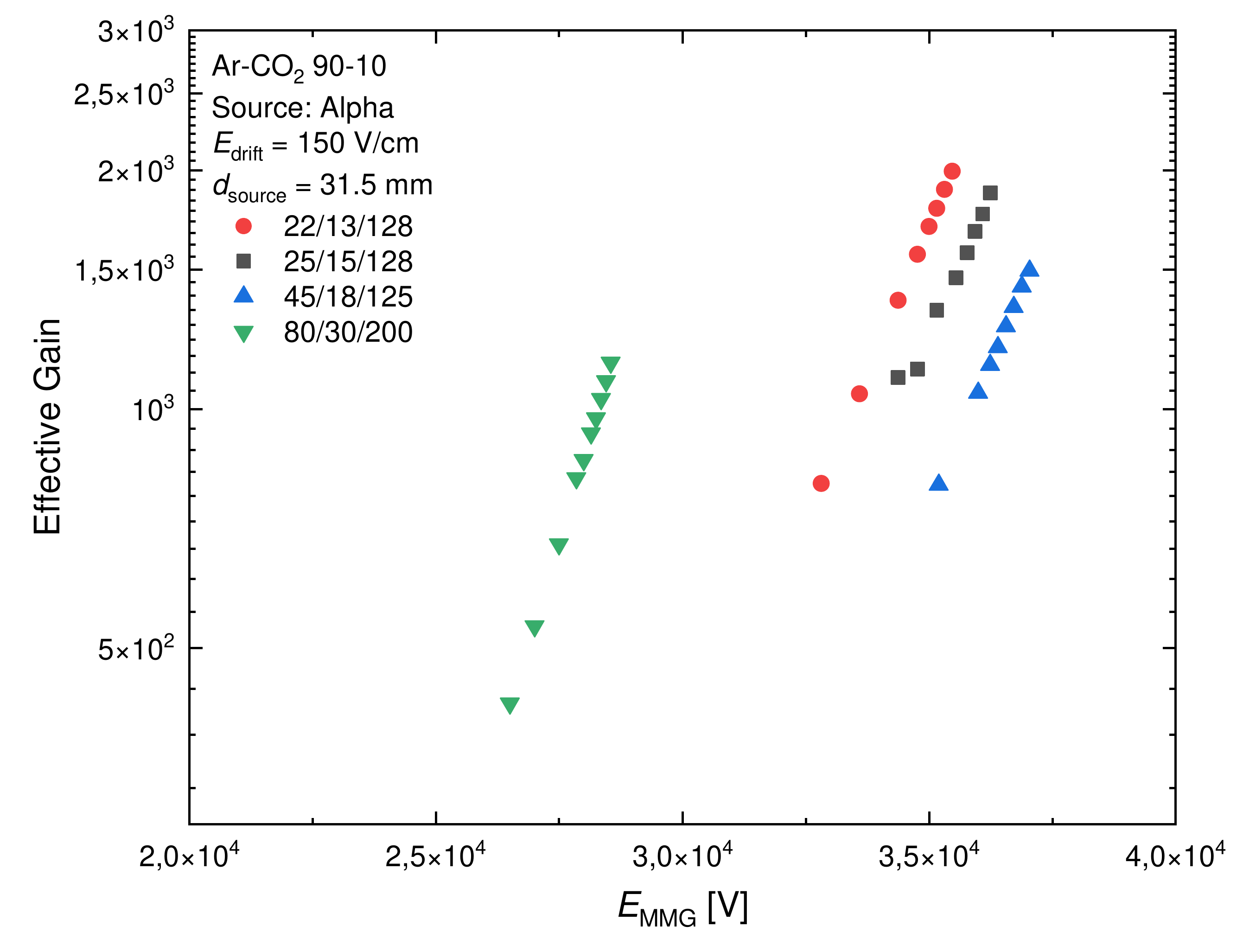}
    \caption{\textit{Left:} discharge probability measured for all four types of Micromegas detectors at 31.5\,mm distance between alpha source and a mesh. \textit{Middle:} effective gain measured as a function of the potential applied to the mesh. \textit{Right:} effective gain plotted as a function of the amplification field. All measurements are performed with the drift field of \SI{150}{\volt\per\centi\meter} for which the same electron transparency in all four meshes can be assumed (see \figref{fig:mmg:transparency}).}
    \label{fig:qcrit:mmgtyp}
\end{figure}

The direct comparison of all four meshes in terms of gain and discharge probability is shown in \figref{fig:qcrit:mmgtyp}. Although the differences between discharge stability measured for different meshes are not large, there is a clear order of discharge curves observed, pointing to the meshes with thin wires and small cell sizes as the more stable ones, similar to~\cite{ALVIGGI2020162359}. We explain this result with the primary charge density hypothesis. With a larger wire pitch, more electrons enter a single mesh cell and are multiplied therein. This could explain the discharge curve scaling with the wire pitch and MMG4 being the less stable structure. This observation would also suggest that a Micromegas mesh cell can be treated as an independent amplification unit, similar to a hole in a GEM foil. 

\begin{figure}[h]
    \centering
    \includegraphics[width=0.33\linewidth]{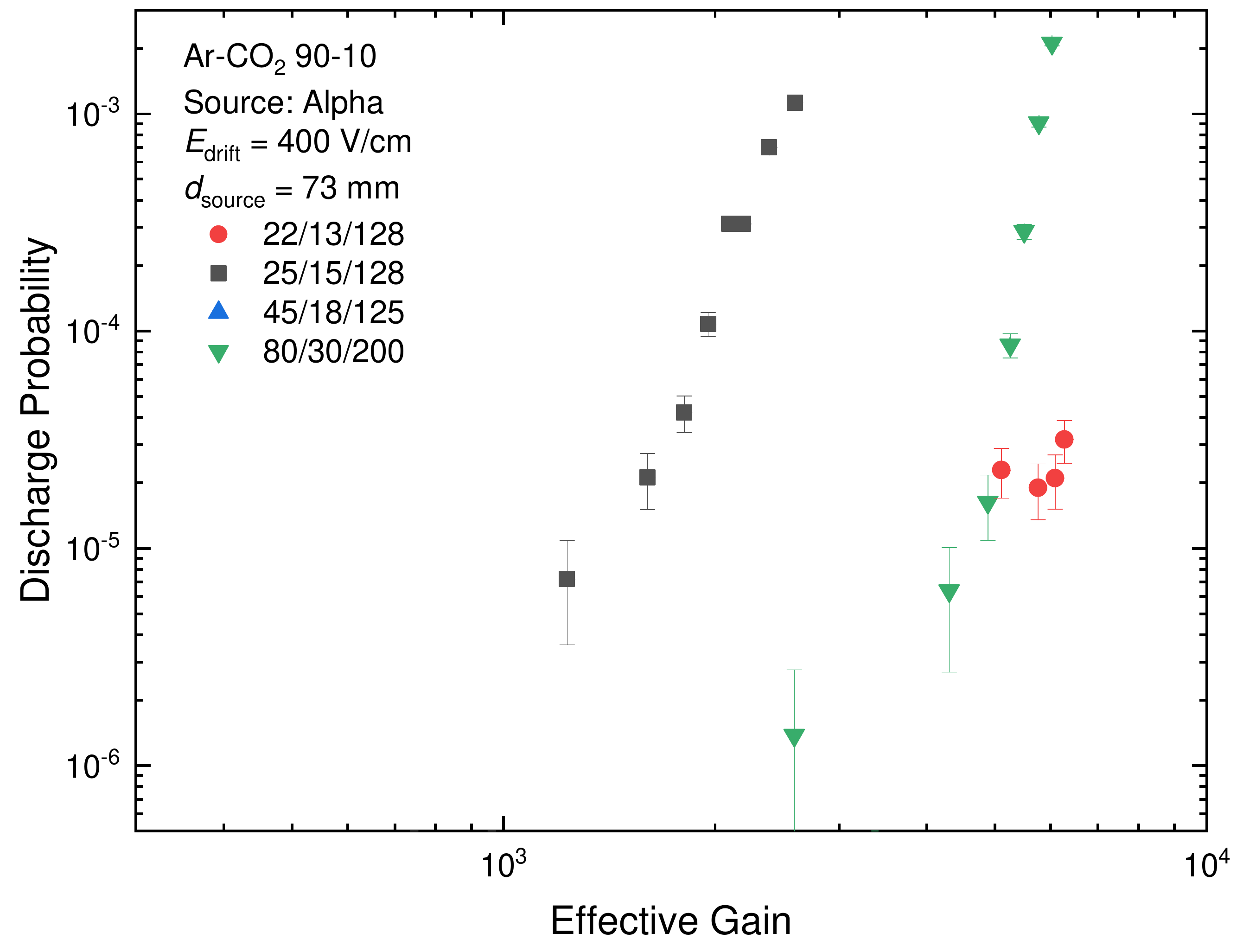}
    \includegraphics[width=0.33\linewidth]{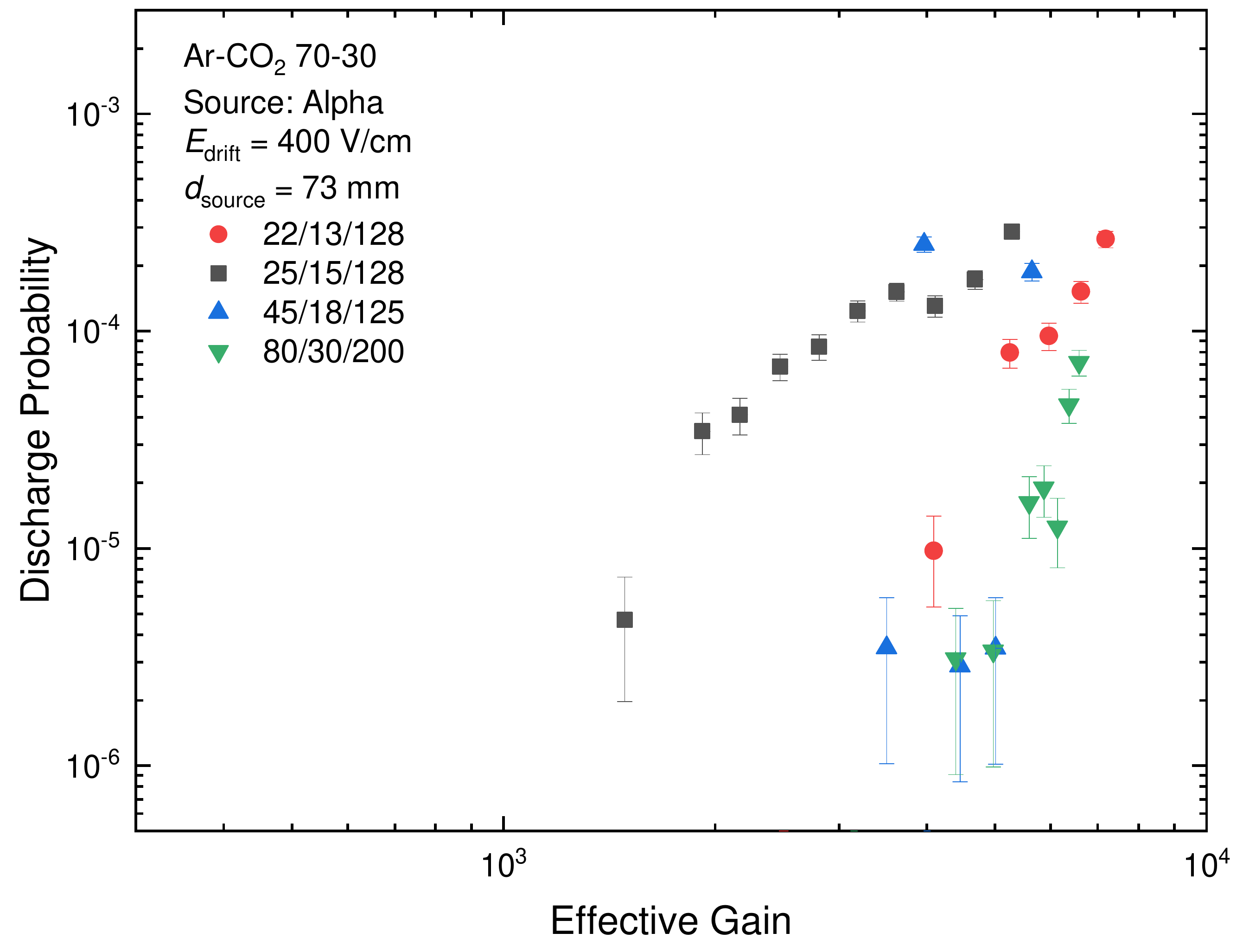}
    \caption{Discharge probability vs. effective gain measured in \ArCOtwo (left) and \ArCOtwoThirty (right) for all four MMG types at 73\,mm distance between the alpha source and the mesh. No discharges have been observed with MMG3 in \ArCOtwo at gain values measurable with the available equipment.}.
    \label{fig:qcrit:bragg}
\end{figure}

Of course, a much simpler interpretation can also be considered. For technological reasons, the MMG4 structure features a larger amplification gap than the other three detectors. This means, that in order to achieve the same gain, larger potential needs to be applied to the structure, which is clearly seen in the gain curves plotted in \figref{fig:qcrit:mmgtyp}. Although the average amplification field in the gap is lower, the peak field around mechanical imperfections or places where two wires of a woven mesh splice may be enhanced, increasing the chance of triggering a discharge. Therefore, a measurement is performed in Ar-based mixtures with the alpha source placed 73\,mm away from the mesh surface, more than the range $r_{\alpha}\approx4.8$\,cm of emitted alpha particles~\cite{MathisGEM}. Thus, all alpha tracks are confined between the source and the mesh and even the highest primary charge densities obtained around the Bragg peak will be reduced during the electron drift toward the amplification structure. As discussed in ~\cite{MathisGEM, LautnerTH}, the discharge probability for these distances drops by several orders of magnitude, close to the background level. The results with the Micromegas detectors are presented in \figref{fig:qcrit:bragg}.

Larger gains are obtained in the situation where no alpha particles penetrate the mesh, and primary charge densities are largely reduced by diffusion. It should be noted, that the measurements at $\dsrc=73$\,mm are available only for the drift field of \SI{400}{\volt\per\centi\meter}, therefore the actual gain of the thin-wire meshes (MMG1 and MMG2) is \SI{\sim20}{\percent} larger than for MMG3 and MMG4, given the electron transparency from \figref{fig:mmg:transparency}. At these voltages, the electric field is enhanced at the hot spots and any kind of imperfections shall be manifested by reduced discharge stability. Indeed, this may explain significantly higher discharge rates obtained with the MMG2 structure. The latter, on the other hand, is the most stable in the measurements at $\dsrc=31.5$\,mm, which points towards the conclusion that the charge density effects exceed those related to the potential hot spots, in this configuration. All the other meshes measured at $\dsrc=73$\,mm show stable behavior up to much higher gains and voltages, without significant dependency on their geometrical parameters. 

This measurement suggests that the dependency presented in \figref{fig:qcrit:mmgtyp} and the one reported in~\cite{ALVIGGI2020162359} may be indeed the consequence of the charge sharing between single mesh cells and not the result of the field uniformity. If confirmed, the observation will give valuable input to the design of future Micromegas-based detectors, especially multi-MMG and hybrid stacks.

\acknowledgments
This work is supported by the Deutsche Forschungsgemeinschaft - Sachbeihilfe [DFG FA 898/5-1].
\linebreak Lukas Lautner acknowledges support from the Wolfgang Gentner Programme of the German Federal Ministry of Education and Research (grant no. 13E18CHA)




\begin{thebibliography}{99}

\bibitem{GEM}
F.~Sauli, \emph{GEM: A new concept for electron amplification in gas detectors}, \href{https://doi.org/10.1016/S0168-9002(96)01172-2}{NIM A 386 (1997) 531}

\bibitem{MMG}
Y.~Giomataris et al., \emph{MICROMEGAS: A High granularity
position sensitive gaseous detector for high particle flux environments}, \href{https://doi.org/10.1016/0168-9002(96)00175-1}{NIM A 376 (1996) 29}

\bibitem{BachmannSpark}
S.~Bachmann et al., \emph{Discharge studies and prevention in the gas electron multiplier (GEM)}, \href{https://doi.org/10.1016/S0168-9002(01)00931-7}{NIM A 479 (2002) 294}

\bibitem{MathisGEM}
P.~Gasik et al., \emph{Charge density as a driving factor of discharge formation in GEM-based detectors}, \href{https://doi.org/10.1016/j.nima.2017.07.042}{NIM A 870 (2017) 116}

\bibitem{LautnerTH}
P.~Gasik et al., \emph{Systematic investigation of critical charge limits in Thick GEMs}, \href{https://doi.org/10.1016/j.nima.2022.167730}{NIM A 1047 (2023) 167730}

\bibitem{Bressan}
A.~Bressan et al., \emph{High rate behavior and discharge limits in micro-pattern detectors}, \href{https://doi.org/10.1016/S0168-9002(98)01317-5}{NIM A 424 (1999) 321}


\bibitem{BAY2002162}
A.~Bay et al., \emph{Study of sparking in Micromegas chambers}, \href{https://doi.org/10.1016/S0168-9002(02)00510-7}{NIM A 488 (2002) 162}

\bibitem{PROCUREUR2010177}
S.~Procureur et al., \emph{A Geant4-based study on the origin of the sparks in a Micromegas detector and estimate of the spark probability with hadron beams}, \href{https://doi.org/10.1016/j.nima.2010.05.024}{NIM A 621 (2010) 177}

\bibitem{ALVIGGI2020162359}
M.~Alviggi et al., \emph{Discharge behaviour of resistive Micromegas}, \href{https://doi.org/10.1016/j.nima.2019.162359}{NIM A 958 (2020) 162359}

\bibitem{Bhattacharya_2020}
D.S.~Bhattacharya et al., \emph{A numerical investigation on the discharges in Micromegas
}, \href{https://doi.org/10.1088/1742-6596/1498/1/012032}{J. Phys. Conf. Ser. 1498 (2020) 012032}

\bibitem{Nikolopoulos_2011}
K.~Nikolopoulos et al., \emph{Electron transparency of a Micromegas mesh}, \href{https://doi.org/10.1088/1748-0221/6/06/P06011}{JINST 6 (2011) P06011}

\bibitem{Bhattacharya_2014}
P.~Bhattacharya et al., \emph{Performance studies of bulk Micromegas of different design parameters}, \href{https://doi.org/10.1088/1748-0221/9/04/C04037}{JINST 9 (2014) C04037}

\bibitem{Krugerphd}
F.~Kruger, \emph{Signal Formation Processes in Micromegas Detectors and Quality Control for large size Detector Construction for the ATLAS New Small Wheel}, PhD thesis, Julius-Maximilians-Universit\"{a}t W\"{u}rzburg, 2017, \href{https://doi.org/10.48550/arXiv.1708.01624}{\tt arXiv:1708.01624 [physics.ins-det]}








\end{thebibliography}

\end{document}